\documentstyle[prc,twocolumn,aps,epsfig]{revtex}

\begin{document}
\draft

\title{\qquad\qquad\qquad\qquad\qquad\qquad\qquad\hfil{NT@UW-00-07,KRL MAP-266}\\  
Chiral Three-Nucleon Forces from
$p$-wave Pion Production}

\author{C. Hanhart$^a$, U. van Kolck$^{b,a}$, and
G.A. Miller$^a$ }

\address{$^a$ Department of Physics,  University of 
Washington, Box 351560, Seattle, WA 98195-1560}

\address{$^b$ Kellogg Radiation Laboratory, 106-38,
California Institute of Technology, Pasadena, CA 91125}

\maketitle

\begin{abstract} 
Production of $p$-wave pions in nucleon-nucleon collisions
is studied according to an improved  power counting that embodies
the constraints of chiral symmetry.
Contributions from the first two non-vanishing orders 
are calculated. 
We find reasonable convergence and agreement with data
for a spin-triplet cross section in $pp\rightarrow pp\pi^0$,
with no free parameters. 
Agreement with existing data for a spin-singlet cross section
in $pp\rightarrow pn\pi^+$
constrains a short-range operator
shown recently to contribute significantly to
the three-nucleon potential.

\end{abstract}

\vspace{0.8cm}
\newcommand{\boldpi}{\mbox{\boldmath $\pi$}}
\newcommand{\boldtau}{\mbox{\boldmath $\tau$}}
\newcommand{\boldT}{\mbox{\boldmath $T$}}
\newcommand{\gaprox}{$ {\raisebox{-.6ex}{{$\stackrel{\textstyle >}{\sim}$}}} $}
\newcommand{\saprox}{$ {\raisebox{-.6ex}{{$\stackrel{\textstyle <}{\sim}$}}} $}

The use of  (approximate) chiral symmetry of QCD to determine the form
of the low-energy
effective Lagrangian has proven to be a powerful aid to the understanding
of strong interaction physics \cite{book}.
It has long been known \cite{TM-Braz,vK94,losalamos} that 
the use of chiral symmetry for pion-nucleon ($\pi N$) scattering
leads to a qualitative understanding of the
pion-ranged part of the three-nucleon force, 
believed  to produce important effects in 
nucleon-deuteron ($Nd$) scattering \cite{gloeckle} and few-nucleon
bound states \cite{vijay}.
Yet, discrepancies between theory and experiment 
(for $A_y$ in $Nd$ scattering \cite{friar}
and for excited levels in bound states \cite{wiringa}) remain which
have been widely attributed to unknown three-nucleon forces.
A novel three-nucleon force, expected on the
basis of power counting arguments,
involves the exchange
of a pion between one nucleon and two others
interacting via short-ranged forces \cite{vK94}.
This force can indeed affect 
$Nd$ scattering at a currently observable level,
and thus potentially resolve the remaining discrepancies
\cite{bochum}. 
It depends on a pion-two-nucleon  interaction of a form
 determined by chiral symmetry, but strength determined by 
parameters, $d_i$ of Eq.~(\ref{la1}),
not fixed by symmetry. We argue here that
the production of $p$-wave pions in nucleon-nucleon 
($NN$) scattering
offers a unique opportunity to determine $d_i$
\cite{biratalk}.

In the last few years, the various $NN\rightarrow NN\pi$
reactions
have been studied both experimentally and theoretically \cite{chrisreview},
with a 
focus  on near-threshold energies. 
The first high-quality data concerned the total
cross section, and
most  theoretical analyses have concentrated
on $\eta \saprox 0.4$, a region dominated by the $Ss$ state.
(Final states are labeled by $Ll$ with $L$ and $l$ being
the relative angular momentum of the nucleon pair and the pion
with respect to the two-nucleon center of mass, respectively; 
$\eta$ is
the maximum pion momentum in units of the pion mass, $m_\pi$). 
Many different mechanisms are expected
at these kinematics: heavy meson exchanges 
\cite{LuR}, (off-shell) pion rescattering 
\cite{HO,Han1}, 
excitations of baryon resonances \cite{Pena}, and  pion emission from
exchanged mesons \cite{CP2b}. 

The pion dynamics are largely controlled by
chiral symmetry constraints, and the  hope that
the use  of Chiral Perturbation Theory ($\chi$PT) would yield
insights led to 
 the use of tree-level
$\chi$PT to calculate the cross sections close to threshold
\cite{CP2,CP1,CP3,HHH,CP4}.
 Ref. \cite{CP2} emphasized that 
the diverse  contributions to the $Ss$ final states
can be ordered 
in powers of $\sqrt{m_\pi/M_{QCD}}$,
where $M_{QCD}\simeq 1$ GeV is the typical QCD mass scale.
The implication of this relatively
large parameter is that loop diagrams
enter  at next-to-leading order in $s$-wave pion production. 
Thus a test of the 
convergence  of the series is hindered.
We shall show  that such difficulties are not present for the case of
$\pi$  production in $p$-waves ($\eta \sim 1$), 
because
the production proceeds through leading-order operators
better determined from other processes.

Our arguments rely on the
use of 
symmetries. 
One may obtain the results of  
 QCD by using 
the most general Lagrangian involving the
low-energy degrees of freedom
(pion $\boldpi$,  nucleon $N$, and delta isobar $\Delta$) which
has the same symmetries as QCD. These are 
approximate chiral symmetry, parity and  time-reversal invariance.
Chiral symmetry plays a crucial role in low-energy processes 
because it demands that, in the chiral limit where the quark masses
go to zero, the pion  interactions contain derivatives, which are
weak at small momenta, $Q$. Because the quark masses are small,
any  non-derivative pion interactions  are also weak.
Although the nucleon mass $m_N$ is not small,
it plays no dynamical role at low energies.
The delta isobar can be excited, but its
mass difference to the nucleon,
$\delta m\equiv m_\Delta - m_N$, is not large.
For processes in which $Q \sim m_\pi$ it is 
convenient to introduce the  ``chiral index'' of an interaction
$\Delta=d+\frac{f}{2}-2$,
where $d$ is the number of small-scale factors, that is, derivatives,
$m_{\pi}$, 
and $\delta m$;
and $f$ is the number of fermion field operators.
Chiral symmetry implies 
that $\Delta\ge 0$ \cite{W90}.
Our interaction Lagrangian is given, using
an appropriate choice of fields, by the expressions \cite{OvK92,ORV94}
\begin{eqnarray}
 {\cal L}^{(0)}_{\rm int} & = & 
-\frac{1}{4 f_{\pi}^{2}}  N^{\dagger}
\boldtau \cdot (\boldpi\times\dot{\boldpi})N +\frac{g_{A}}{2 f_{\pi}} 
         N^{\dagger}(\boldtau\cdot\vec{\sigma}\cdot\vec{\nabla}\boldpi)N
                                               \nonumber \\
    &   & 
          +\frac{h_{A}}{2 f_{\pi}}[N^{\dagger}(\boldT\cdot
          \vec{S}\cdot\vec{\nabla}\boldpi)\Delta ] +\cdots 
\label{la0}
\end{eqnarray} 
and \cite{vK94}
\begin{eqnarray}
&& {\cal L}_{\rm int}^{(1)} 
        =\frac{i}{8m_{N}f_{\pi}^{2}} 
        N^{\dagger}\boldtau\cdot
        (\boldpi\times\vec{\nabla}\boldpi)\cdot\vec{\nabla}N 
         -\frac{c_3}{f_{\pi}^{2}}N^{\dagger}
        (\vec{\nabla}\boldpi)^{2}N                           \nonumber \\
  &   & -N^{\dagger}
   \frac{\bar c_4}{2f_{\pi}^{2}}
  \vec{\sigma}\cdot 
        \vec{\nabla}{\boldpi} \times\vec{\nabla}
        \boldpi\cdot\boldtau N
        -\frac{ig_{A}}{4 m_{N} f_{\pi}}N^{\dagger}\boldtau\vec{\sigma}
          \cdot\dot{\boldpi}
        \vec{\nabla}N    \nonumber   \\  
                                                               \nonumber \\
  &   &        -\frac{h_{A}}{
        2 m_{N} f_{\pi}}[
        iN^{\dagger}\boldT\cdot\dot{\boldpi}\vec{S}\cdot\vec{\nabla}
        \Delta]                      
        -\frac{d_1}{f_{\pi}} 
        N^{\dagger}\boldtau\cdot\vec{\sigma}\cdot\vec{\nabla}\boldpi N\,
        N^{\dagger}N \nonumber \\
&   &        -\frac{d_2}{2 f_{\pi}}  
        \vec{\nabla}\boldpi \times  
        N^{\dagger}\vec{\sigma}\boldtau N\;\cdot
        N^{\dagger}\vec{\sigma}\boldtau N 
        +\cdots,
        \label{la1}
\end{eqnarray}
\noindent
where $\bar c_4 = c_4+\frac{1}{4m_N}$.
The terms denoted by $\cdots$ include Hermitian conjugates,
$s$-wave $\pi N$ scattering terms and
terms of higher powers in pion fields. 
Our principle aim is 
 to determine the parameters $d_{i}={\cal O}(1/f_\pi^2 M_{QCD})$,
 which determine the desired three-nucleon force.
Here we fix signs by taking $g_A=+1.26$ in the chiral limit.
The $c_i$ have been determined from $\pi N$ scattering at
tree level 
($c_3^{(tree)}  = -3.90 \ \mbox{GeV}^{-1}$
and 
$c_4^{(tree)} =  2.25 \ \mbox{GeV}^{-1}$ \cite{bkmppn}) 
as well as to one-loop order 
($c_3^{(loop)}  = -5.29 \ \mbox{GeV}^{-1}$
and $c_4^{(loop)} = 3.63 \ \mbox{GeV}^{-1}$ \cite{bkmci}).
(Since we treat the delta isobar explicitly, we 
subtract its contribution
from these values of $c_i$ \cite{askbira}.)
As will be established below, up to next-to-leading
order, the $d_i$, which  support only  $S \to Sp$ transitions, 
 are the only undetermined parameters in $p$-wave pion production.

The next step is to extend 
 the power counting of Ref. \cite{CP2}
to the region $\eta \sim 1$, where the outgoing pion has 
energy $\omega={\cal O}(m_\pi)$
{\it and} momentum $|\vec{q}| ={\cal O}(m_\pi)$,
and the two nucleons in the final state 
have momentum 
$|\vec{p'}|={\cal O}(m_\pi)$
and total energy
$p'^0={\cal O}(m_\pi^2/m_N)$.
The unique difficulty of  using $\chi$PT for pion production
is that the entire pion energy is supplied
by the relatively large momentum of the initial nucleons,
$|\vec{p}|={\cal O}(\sqrt{m_N m_\pi})$.
Note that the 
non-relativistic approximation 
holds, as $p^4/8m_N^3 \sim m_\pi^2/m_N \ll m_\pi \sim p^2/2m_N$.

The scales of momenta and energy are
not the same, so it is simpler to count powers of the small scales
in time-ordered perturbation theory. Equivalently, one first integrates 
 over the time  component of
loop momenta in covariant diagrams.
In this case, an intermediate state is associated with
an energy denominator $1/E$, a loop with a $Q^3/(4\pi )^2$,
a spatial (time) derivative with $Q$ ($E$),
and a virtual pion vertex with $1/E^{1/2}$
from wave function normalization. 
For $N$, $E\sim Q^2/m_N$, for  
$\Delta$, $E\sim Q^2/m_N +\delta  m$,
and for $\boldpi$,  $E\sim \sqrt{Q^2+m_\pi^2}$.

 Final-state interactions (FSI)
are those which occur after the emission of the real pion.
In this case,  the nucleons have typical $Q \sim m_\pi$. 
The energies of intermediate states containing a
                                $\boldpi$ or $\Delta$ 
can be $E\sim m_\pi$, but 
otherwise $E \sim m_\pi^2/m_N$.
The sum of ``irreducible'' 
sub-diagrams where all energies are ${\cal O}(m_\pi)$
is by definition the $NN$
potential, which is then amenable to a $\chi$PT expansion.
The sum of ``reducible'' sub-diagrams
 produces the final-state wave function
$|\psi_f\rangle$.

In contrast,
all intermediate states occurring before the radiation of the
real pion
are characterized by loop momenta
$ \sim\sqrt{m_N m_\pi}$.
For these kinematics we find that 
any additional loop requires at least 
{\it i)} one more interaction ---pion exchange or shorter range---
with an associated factor no larger than $1/f_\pi^2$;
{\it ii)} a volume integral with an associated factor of
$(\sqrt{m_\pi m_N})^3/(4\pi)^2$;
and 
{\it iii)} an additional time slice.
If the additional time slice cuts a pion line,
a factor of $1/\sqrt{m_\pi m_N}$ comes in, and the
overall extra loop factor is at least
$
\frac{1}{f_\pi^2}
\frac{(\sqrt{m_\pi m_N})^3}{(4\pi)^2}
\frac{1}{\sqrt{m_\pi m_N}}
=\frac{m_\pi}{m_N},
$
that is, a suppression by two powers of the
expansion parameter. If the additional time slice does not cut a pion line,
a factor of $1/m_\pi$ appears, and there is 
 a relative enhancement of
$\sqrt{m_N/m_\pi}$.
Integrals over two-nucleon states typically also
have enhancements by factors of $\pi$ from the unitarity cut.
Thus we  resum  those diagrams that differ
by the addition of interactions between the initial
nucleons (ISI), and
the effects are contained in an initial state wave function
$|\psi_i\rangle$.

These considerations yield a pion production
amplitude 
$
T= \langle\psi_f|K|\psi_i\rangle.
$ 
Both the kernel  $K$ and           $|\psi_{i,f}\rangle$ 
can be obtained from the chiral expansion, but
 the currently available  $|\psi_i\rangle$ 
 do not yield an accurate fit to
the measured
 $NN$ scattering phase shifts.  Therefore we use 
a phenomenological  coupled-channel ($NN$, $N\Delta$, $\Delta\Delta$) model,
CCF of Ref.~\cite{HHJ}, fitted to $NN$ scattering.

\begin{figure}[tb]
\vspace{2.9cm}
\includegraphics{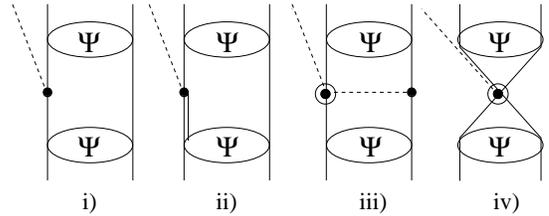}
\caption{ Lowest-order contributions to $p$-wave 
production.
Diagrams at 
${\cal O}(1)$ are (i,ii),
and of ${\cal O}(m_\pi/m_N)$ are (iii,iv). 
A solid (dashed) line
denotes a nucleon (pion),
and a double line a $\Delta$.
Interactions from ${\cal L}^{(0)}$ (${\cal L}^{(1)}$)
are denoted by a dot (circled dot).
Diagrams with a 
$\Delta$ in the final state are also included. 
}
\label{pi0con} 
\end{figure}

The leading contributions to $p$-wave production
are displayed in Fig. \ref{pi0con}.
At lowest order (${\cal O}(1)$, apart from overall factors)
there are contributions from the direct
production off the nucleon and off the delta,
where all vertices are from ${\cal L}^{(0)}$ (Fig. \ref{pi0con}i, ii). 
At next-to-leading non-vanishing order 
(${\cal O}(m_\pi/m_N)$)
there are four types of contributions.
First, there is a recoil 
correction to the direct production.
Second, there are rescattering diagrams
that proceed through the seagull vertices in ${\cal L}^{(1)}$
proportional to $1/4 f_{\pi}^{2}$ (Galilean correction
to the Weinberg-Tomozawa term),
$c_3$, and $c_4$ (Fig. \ref{pi0con}iii).
Third, there is 
a rescattering through the Weinberg-Tomozawa term, where the primary
production vertex is proportional to the external pion momentum.
Fourth, there are short-range $\pi (N^\dagger N)^2$
interactions proportional to $d_1$ and $d_2$ (Fig. \ref{pi0con}iv).
Diagram  iv) and most of the rescattering
diagrams contribute to charged-pion production only.

With our theory in place, we consider the available 
 pion production database. This has been enriched recently 
by very accurate determinations of 
spin observables at $0.5 \saprox \eta \saprox 1$ for 
$pp\rightarrow pp\pi^0$ \cite{meyer}, 
$pp\rightarrow pn\pi^+$ \cite{daehnick},
and $pp\rightarrow d\pi^+$ \cite{usualsuspects}.
It is useful to describe  the total cross section in terms of
components $^{2S+1}\sigma_m$, where $S$ is the initial $NN$ spin
with projection $m$ along the direction of the incoming momentum.
The  $^{2S+1}\sigma_m$ can be expressed as 
linear combinations of the total cross section and the
double polarization observables 
$\Delta \sigma_T$ and $\Delta \sigma_L$ \cite{meyer}. 

In order to test convergence for the $p$-wave production,
we need an observable where the lowest contributing partial wave
is $p$ and the initial and final nucleons are 
not both in $S$ states. 
Such an observable exists, namely the $^3\sigma_1$ cross section
in neutral-pion production with $Pp$ as the lowest partial waves contributing.
While  the ratios between
double polarization observables and the total cross section,
$\Delta \sigma_T/\sigma_{\rm tot}$ and 
$\Delta \sigma_L/\sigma_{\rm tot}$, have recently been 
accurately measured at IUCF \cite{meyer},
the total cross section is known to a much lesser
accuracy (see the  compilation of 
Ref. \cite{DATA}). To determine the error
of the total cross section, we simply take the total 
spread of the data as the error band.
We defer a more detailed 
analysis until it can benefit from
the soon-to-be-available \cite{jozef}
much better data.

In $p$-wave production the lowest-order 
loop contribution enters one order higher, at 
${\cal O}((m_\pi/m_N)^\frac{3}{2})$, than the rescattering terms of
Fig.~\ref{pi0con}, and are ignored.
Besides the coupling constants of the pion to the baryon fields, the
only parameter that enters is $c_3$. 
We will use both values given above to get
an estimate of loop effects on the final result.

\begin{figure}[tb]
\vspace{6cm}
\includegraphics{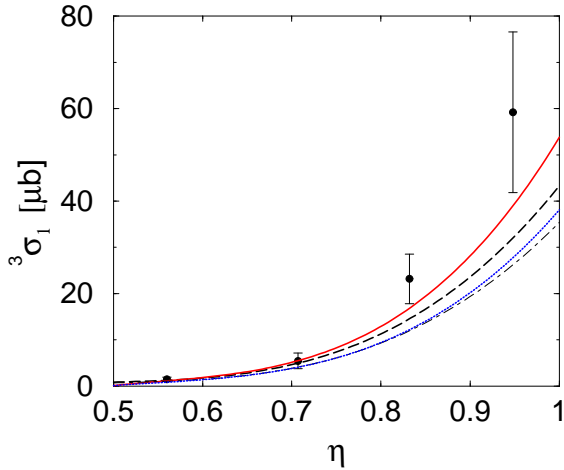}
\caption{Chiral perturbation theory predictions for $^3\sigma_1$ 
in the reaction $pp \to pp\pi^0$. 
Lowest order (long dashed line), 
lowest order plus recoil contribution (dot-dashed line), 
and next-to-leading order using $c_3^{(loop)}$ (solid line) 
and $c_3^{(tree)}$ (dotted line) are shown.
Data are from Refs. \protect\cite{meyer,DATA}.}
\label{threesigone_pi0} 
\end{figure}

The {\it predictions} of $\chi$PT are compared to 
the data in Fig. \ref{threesigone_pi0}. 
Up to values of $\eta \simeq 0.7$ the data
is well described. 
Deviations at higher energies might be due to higher partial
waves entering,
and/or to higher-order $p$-wave contributions.
In any case, we see that sub-leading corrections are smaller
than leading contributions throughout the range
$\eta \saprox 1$.

We next consider the amplitude for the $^1S_0\to(^3S_1-^3D_1)p$
transition, denoted $a_0$, which has recently been extracted
from the reaction $pp\to pn\pi^+\cite{Flammang}$.
The loop corrections are again expected to be
small, but the 
 number of rescattering diagrams is larger, since  isospin-odd operators 
 (the recoil correction to the Weinberg-Tomozawa
term as well as the $c_4$ term  of  Eq.~(\ref{la1})) enter. 
The striking feature of $a_0$ is that
interactions proportional to the $d_i$'s also contribute. 
Because there seems to be reasonable convergence
in the $p$ waves, we assume that they can be reliably computed and
that we can attribute any deviation between theory and experiment to
the effects of the terms involving 
the coefficients $d_{i}$. 
The  contact
interactions enter as the  linear combination 
 $d_1 + 4d_2$. Thus
there is one unknown parameter to be fixed by the data. 
On the basis of dimensional analysis 
we expect 
$
d\equiv \frac{1}{5} (d_1 + 4d_2)= \frac{\delta}{f_\pi^2 M_{QCD}}
$ 
with
$\delta= {\cal O}(1)$.

\begin{figure}[tb]
\vspace{5.2cm}
\includegraphics{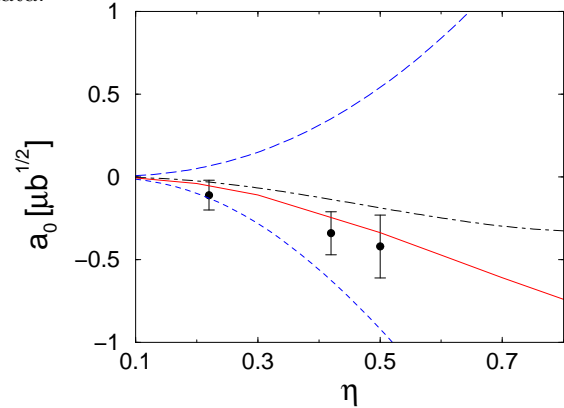}
\caption{ $a_0$
of  $pp \to np\pi^+$ in chiral perturbation theory.  
The different lines correspond to 
values of the parameter related to the three-nucleon force:
$\delta=1$ (long dashed line).
$\delta=0$  (dot-dashed line),
$\delta=-0.2$ (solid line), and
$\delta=-1$ (short dashed line).
Data are  from Ref. \protect\cite{Flammang}. }
\label{a0pn} 
\end{figure}

Our result for $a_0$ is shown in Fig.~\ref{a0pn}.
We find a destructive interference 
between direct nucleon and delta contributions
that makes  $a_0$ small and more sensitive to
sub-leading terms.
For the $c_i$ parameters we employ the values extracted
from the tree level fit to $\pi N$ scattering ($c_i^{(tree)}$). 
We use dipole form factors;
to make contact with Ref. \cite{bochum}, we employ
cutoff parameters $\Lambda = 1$ GeV
for diagrams containing pion exchange
and $\Lambda = m_\omega$ 
for the contact interactions.
The result for $\delta =0$ is
not in disagreement with data, whereas
a value of $\delta=1$ leads to a serious disagreement
with experiment.
In Ref. \cite{bochum} $\delta=-0.2$ was shown to yield 
an important contribution
to $A_y$ in $Nd$ scattering at energies of a few MeV.
Using  $\delta=-0.2$ here is also 
consistent with the pion production data.

In contrast to $^3\sigma_1$,
the result for $a_0$ is quite sensitive to the cutoff 
parameter used in the rescattering contribution, because
 the momentum range scanned by the $c_4$ term is quite large.
For example, 
our results for $a_0$ can vary up to a factor of 2
if the corresponding cutoff parameter is 
increased to 2 GeV.
The cutoff sensitivity is not a serious difficulty because
it also occurs in calculations of 
three-nucleon forces. 
From the viewpoint of an effective field theory
this can be simply understood:
the large momentum
pieces of the loop integrals involved in the evaluation
of the $c_4$ contribution
can be absorbed by a counterterm, namely $d_2$.
Thus, the cutoff dependence
of $c_4$ directly translates into a scale dependence of $d_2$.
A reasonable phenomenological estimate should follow from using
the same cutoff and parameter set in both calculations.
On the experimental side, it is clear from Fig. \ref{a0pn}
that a reduction of
the uncertainty in the data would allow a
stronger constraint on $\delta$.
We find this a strong motivation to the continuation
of the existing program on pion production.

We have shown that there is convergence in $p$-wave
pion production, and that data on this reaction can 
be used to extract information about the three-nucleon
force.
It is clear that more accurate data would be very useful.
In particular, the parameter $d$ could be extracted and
the calculation of Ref. \cite{bochum}
repeated to predict three-nucleon observables.
We find it very gratifying that chiral symmetry
provides a direct connection
between pion production at energies $\sim 350$ MeV (IUCF)
and $Nd$ scattering at energies  
$\sim 10$ MeV (Madison, TUNL).

We would like to thank Jim Friar, Brad Keister, and Daniel Phillips
 for useful discussions.
This research was supported in part by the U.S. DOE, the 
NSF, 
and the 
Humboldt
Foundation.

\end{document}